# Flux quanta driven by high-density currents in low-impurity V$_3$Si and LuNi$_2$B$_2$C: free flux flow and fluxon-core size effect

A. A. Gapud, S. Moraes, R. P. Khadka, P. Favreau, and C. Henderson

*University of South Alabama, Department of Physics, 307 University Blvd N, Mobile, AL 36688*

P. C. Canfield and V. G. Kogan

*Ames Laboratory, 111 TASF, Ames, IA 50011-3020*

A. P. Reyes and L. L. Lumata

*National High Magnetic Field Laboratory, 1800 E Paul Dirac Drive, Tallahassee, FL 32310-3706*

D. K. Christen

*Oak Ridge National Laboratory, 1 Bethel Valley Rd, Oak Ridge, TN 37831-6092*

J. R. Thompson

*U. of Tennessee, Department of Physics and Astronomy, Knoxville, TN 37996-1200 and*
*Oak Ridge National Laboratory, 1 Bethel Valley Rd, Oak Ridge, TN 37831-6092*

## Abstract

High density direct currents (DC) are used to drive flux quanta via the Lorentz force towards a highly ordered "free flux flow" (FFF) dynamic state, made possible by the weak-pinning environment of high-quality, single-crystal samples of two low-$T_c$ superconducting compounds, V$_3$Si and LuNi$_2$B$_2$C. We report the effect of the magnetic field-dependent fluxon core size on flux flow resistivity $\rho_f$. Much progress has been made in minimizing the technical challenges associated with the use of high currents. Attainment of a FFF phase is indicated by the saturation at highest currents of flux-flow dissipation levels that are well below the normal state resistance and have field-dependent values. The field dependence of the corresponding $\rho_f$ is shown to be consistent with a prediction based on a model for the decrease of fluxon core size at higher fields in weak-coupling BCS $s$-wave materials.

# I. INTRODUCTION

A current issue in the flux dynamics of mixed state superconductors is the nature of the finite-sized, non-superconducting cores of these magnetic flux quanta[1,2] – henceforth termed "fluxons." While most analyses and applications do not consider the details of the finite size and anisotropic shape of the fluxon core, there are important effects, especially at lower temperatures and higher magnetic fields. This arises mainly from the unique electronic structure of the non-superconducting material within the core, which is not yet fully understood. Fluxon core structure and shape greatly affect the way the vortices interact. For example, new types of phase transitions from one type of flux-lattice symmetry to another have been predicted and discovered as a result of nonlocal electrodynamic effects and interaction with the physical crystal lattice.[3,4] The internal structure of the core also determines the viscous force against which fluxons move under current-driven Lorentz forces, which also affects how phase transitions occur and, in turn, determines the useful current-carrying capacity of a superconductor.[1]

These issues have been explored in previous works on various materials, using methods such as: (1) small-angle neutron scattering (SANS) which first revealed the lattice-symmetry transitions[5] and has subsequently shown a number of structural features in the flux lattice[3,4,6-11], (2) scanning tunneling microscopy[12] which has revealed the existence of internal electronic states within the core and also confirmed the lattice-symmetry transitions seen by SANS[5], (3) magnetization measurements[13] which have revealed a field dependence to the fluxon core size and (4) muon spin rotation spectroscopy[2] which have not only confirmed the field dependence of the core size but also correlated them with the observed lattice-symmetry transitions. However, complementary DC transport measurements, proven effective in probing the behavior of fluxons, have not yet been attempted towards the specific question of fluxon core size effects – even though the experimental signature of these phenomena has already been predicted.

The application of direct current is arguably the most direct way of providing the Lorentz force necessary to drive the motion of fluxons (so-called "flux flow"), while at the same time quantifying the dissipation voltage per unit current associated with this motion, i.e., the flux flow resistance. Free flux flow resistivity $\rho_f$ refers to the ohmic-like dissipation process in which vortices move in a so-called free-flux-flow (FFF) regime wherein the viscous drag on the moving, interacting fluxons greatly exceeds any residual pinning forces present; this leads to a highly ordered movement of fluxons.[14] The dependence of $\rho_f$ on magnetic field $H$ ($\approx$ flux density $B$ in cgs units) is traditionally modeled using the linear, Bardeen-Stephen (BS) relation at a fixed temperature[15]:

$$\rho_f = \rho_N \frac{H}{H_{c2}} \qquad (1)$$

where $\rho_N$ is the extrapolated normal-state resistivity and $H_{c2}$ is the upper critical field for that temperature. For clean, weak-coupling BCS $s$-wave materials[16], Kogan and Zhelezina (KZ) have predicted a deviation from this expression due to a field-dependent core size, $\xi(H) = \xi^*\xi_{c2}$, where $\xi_{c2} = \sqrt{\phi_o/(2\pi\mu_o H_{c2})}$ is the usual coherence length, commonly assumed to be field-independent. For the BS expression (1), substituting $\xi_{c2} \rightarrow \xi(H)$ yields the modified form[13],

$$\frac{\rho_f}{\rho_N} = \frac{H}{H_{c2}} \rightarrow H\frac{2\pi\mu_o\xi^2(H)}{\phi_o} = h\xi^{*2} \qquad (2)$$

where $h = H/H_{c2}$. In the field-dependent KZ picture, for reduced temperature $t = T/T_c < 0.5$ and $h > 0.6$, the quantity $\xi^* = \xi(H)/\xi_{c2}$ is found to be independent of material parameters[16] for relatively "clean" (weakly scattering) materials with scattering parameter $\lambda = hv/2\pi kT_c l \leq 1.0$. Here, $v$ is the average Fermi velocity and $l$ is the electronic mean free path. Under these conditions, all $\xi^*(h,t)$ collapse onto the $\xi^*(h, t = 0)$ curve. For higher $(h,t)$, $\xi^*(h,t)$ tend towards $\xi^*(h, t = 0)$ only as $\lambda \rightarrow 0$. By contrast, increasing $\lambda$ brings $\xi^*(h,t)$ towards unity (constant). More interestingly, raising the reduced temperature $t$ has the same effect[16] as raising $\lambda$. Using the numeric solutions[16] for $\xi^*(h)$ for $h > 0.15$ and low $\lambda$, curves have been generated for $\rho_f/\rho_n$ versus $h$ for $t = 0$ and $t = 0.5$; in these numerical results (shown in Fig. 4), the curve for higher $t$ indeed lies closer to BSFF ($\xi^* = 1$). Towards confirming these predictions, this study measures the normalized $\rho_f$ for samples with low $\lambda$, at the vaporization temperature of liquid helium which is almost halfway between $t = 0$ and $t = 0.5$, and for the largest possible range of magnetic fields $h > 0.15$. The results of this investigation reveal that, indeed, the field dependence of $\rho_f$ is consistent with the KZ prediction.

## II. EXPERIMENTAL DETAILS

A key technical challenge in examining the flux medium is the ability to achieve a "textbook" flux lattice: one that is relatively free of any pinning, thermal fluctuations, or distortions due to electronic anisotropy (so that the flux lines are less like barely connected "pancake vortices" and more like uniform "rods"). This favors conventional superconductors that are relatively isotropic and have critical temperatures ($T_c$) low enough to minimize thermally induced effects. The latter is most conveniently quantified using the Ginzburg number, $Gi$, defined as[17]:

$$Gi = \frac{\gamma^2}{2}\left(\frac{kT_c}{4\pi\mu_o H_c^2(0)\xi_o^3(0)}\right)^2 \qquad (3)$$

where $H_c$ is the thermodynamic critical field, $\xi_o$ is the coherence length, and $\gamma^2$ quantifies the effective supercarrier mass anisotropy. For this reason, samples of $V_3Si$ and $LuNi_2B_2C$ were used, where $\gamma^2 \approx 1$, and $Gi \sim 10^{-7}$. (By contrast, high-$T_c$ cuprate superconductors, with $\gamma^2 > 25$, have $Gi > 10^{-2}$.) In addition, this added benefit of very low anisotropy encourages more three-dimensional flux motion. Another requirement, to minimize pinning, is quality materials that contain very few defects, something difficult to achieve. The quality of the $V_3Si$ and $LuNi_2B_2C$ samples in this study has already been demonstrated in other, previous work.[13, 18-20] In addition, a measurement of the residual resistivity ratio (RRR), for which a sample could be considered sufficiently "clean" at values of ~10, yields values exceeding 35, as shown in Fig. 1. The scattering parameter $\lambda$ previously defined was also determined: For the present $V_3Si$ material, this is estimated at 0.38. (Here we use root-mean-square $v = (2.94 \times 10^{14}\ cm^2/s^2)^{1/2}$ from band-structure calculations[3], with the value $l = 32$ nm from a previous study[4] on the same sample.) For the $LuNi_2B_2C$ sample, $\lambda$ is estimated at 0.25. (Using $\rho l = 0.42 \times 10^{-5}\ \mu\Omega\ cm^2$, also from band-structure calculations[21], and $\rho(T_c) = 1.0\ \mu\Omega\ cm$, one obtains $l = 42$ nm; also, root-mean-square $v_a = (1.87 \times 10^{14}\ cm^2/s^2)^{1/2}$ – since the sample is a single crystal.[22])

Experimentally, driving the fluxons toward a FFF phase is done by pulsing the current through the sample in a four-terminal strip geometry. Pulse widths were between 17 ms and 50 ms duration, and in opposite polarity in order to eliminate thermal voltage offsets, and voltage measurements were carefully timed so that sampling occurred at the center of each pulse. Another technical challenge is the necessity of applying high current densities through bulk samples, which required currents exceeding 50 A (provided by a 100-A Valhalla current calibrator). In addition to limiting the duty cycle by pulsing the current, a very low-resistance sample circuit was constructed by using thick current leads – the thickest are gauge "0000" wires used to connect the current source to the cryostat probe – and by minimizing the contact resistance between sample and current leads. The latter was done using ultrasonic soldering, by which oxides on the contact surfaces are simultaneously lifted off by ultrasonic cavitation to encourage wetting, which was especially necessary due to the difficulty of bonding with bulk samples. In this way, contact resistance was limited to the level of micro-ohms. Another consideration is the fact that high current is applied while the sample is in a dissipative state, raising the possibility of sample heating; to minimize this, the sample was kept submerged at all times in liquid helium at 4.2 K (as also closely monitored by weakly field-dependent Cernox temperature sensors mounted with the sample). Because of all the above measures, dissipation in the mixed state could be limited to levels well below that causing boiling of the cryogenic film around the sample.[23]

## III. RESULTS AND DISCUSSION

Results for electrical transport in the superconductive state are presented in Fig. 2, as plots of the voltage-current ratio ($V/I$) versus current. The semi-log plots indicate that indeed the dissipation levels saturate at highest currents, and that the saturation levels are well below the level for $R_n$ (indicated by the dashed line) and are field-dependent. In the figure, the normal-state-transition resistances $R_n$ at 4.2 K are the saturation levels measured at the corresponding $H_{c2}$ – which is defined where $J_c(H)$ drops below 1 A/cm$^2$ (Later, important magnetoresistive effects will be described.). To obtain the ratio $\rho_f/\rho_n = R_f/R_n$, one must determine the flux flow resistance $R_f$, which is the level towards which the ($V/I$) curves saturate. These saturation levels were obtained via a best fit of the ($V/I$) versus $I$ data to the empirical asymptotic form, $R_f(1 - I_c/I)^\alpha$, yielding the typical curves also shown in Fig. 2.

A frequent signature of weak-pinning systems is the presence of a small window of re-entrant pinning that overcomes the elasticity of the flux medium, which occurs just below critical field $H_{c2}$ ($T_{c2}$). This is manifested as an anomalous "peak" in the otherwise monotonic field dependence of critical current density, $J_c(H)$ – the so-called $J_c$ "peak effect"[24] – shown in the insets of Fig. 2 for the present materials. Being an indicator of more effective pinning, this $J_c$ "peak" has been known to disrupt the formation of a FFF phase[19], and thus the "onset" field (the lower-field bound of the $J_c$ peak) serves as a practical upper boundary for the FFF phase. As expected, only below this onset are voltage-current (VI) curves seen to: (i) saturate to constant levels at highest currents at resistivity levels that are (ii) below that for the normal state $R_n$ and are (iii) field-dependent. In this study, an upper bound was determined to be 3.0 T for LuNi$_2$B$_2$C and 14.0 T for V$_3$Si, as indicated by the vertical line and the labeled arrow. The transport data in the main panels of Fig. 2 lie under these respective field boundaries.

The resulting experimental field dependencies of $\rho_f/\rho_n$ are plotted in Fig. 4, together with the theoretical predictions of Bardeen-Stephen (BSFF) and Kogan-Zhelezina (KZ). Most remarkable is that the LuNi$_2$B$_2$C data are consistent with the KZ curves. As previously described, KZ predicts that elevating the temperature above zero has the same effect as increasing the scattering as quantified by the parameter $\lambda$: a weakening of the field dependence of the fluxon core size, which would be manifested here by a curve shifted closer towards the BSFF. If one considers the sample as clean at $\lambda = 0.25$ (<1.0), then the shift is more likely due to $t$ being halfway between 0 and 0.5 – just as the data lie almost halfway between the KZ curves for $t = 0$ and 0.5. Thus LuNi$_2$B$_2$C data are consistent with the KZ-predicted effect of varying $t$.

Analysis of the V$_3$Si data is more complex, due to substantial magnetoresistive effects on the values of "$R_n$" as well as the presence of a Martensitic transformation at around T ~ 21 K, both of which had been studied by Zotos et al.[25]. Qualitatively, this effect inflates the value of $\rho_n$ – i. e., the resistivity of the normal-state fluxon cores – to a level dependent on field $H$. Indeed, in the $\rho$-$T$ curve of V$_3$Si in Fig. 1, the measured value of $\rho_n = 1.56$ μΩ cm, open circle at 4.2 K, $H = H_{c2} = 18.3$ T, is actually higher than $\rho(T = T_c, H = 0) = 1.39$ μΩcm. By comparison, for LuNi$_2$B$_2$C, the measured value of $\rho_n = 1.8$ μΩ cm (square symbol), lies below the value $\rho(T=T_c) = 2.0$ μΩ cm, consistent with previous studies[26] verifying negligible magnetoresistance in this compound. (The dashed-line curve is not a fit, but a guide to the eye.) In order to obtain reasonable values of magnetoresistivity $\rho_n(H)$ at 4.2 K for fields within the FFF regime, $R(T,H)$ curves in the normal state were obtained for fields up to 9 T on the same sample – as shown in Fig. 3. By then extrapolating $T^2$ fits, resistivities at 4.2 K were obtained and fit – along with the data point of $\rho_n$ ($T = 4.2$ K, $H = H_{c2}$) = 1.56 μΩ – to the Kohler Rule form, $\Delta\rho/\rho_o = A (H/\rho_o)^\beta$, with $A = 0.0151$ (μΩ-cm/T)$^{-2}$ and $\beta = 1.34$; this is shown in the Fig. 3 inset. Here, $\rho_o = \rho(T = 4.2$ K, $H = 0)$. The formula was then used to interpolate all values $\rho_n(H)$ used for $\rho_f/\rho_n$ in Fig. 4 which are indicated by solid triangles. For contrast, one finds that ignoring the magnetoresistive effects and taking $\rho_n$ ($T = 4.2$ K, $H = H_{c2}$) = 1.56 μΩ would have yielded the lower curve shown by open symbols whose slope would have been inconsistent with an approach of $\rho_f/\rho_n$ towards 1 at $h = 1$. (Interestingly, the Kohler Rule form obtained on this material is different from that obtained previously by Zotos et al., a fact which has led to another in-depth inquiry into the magnetoresistivity of these particular samples of V$_3$Si which is currently ongoing.)

Qualitatively, the resulting $V_3Si$ results are quite similar to those for $LuNi_2B_2C$, with both deviating significantly from BSFF predictions and showing consistency with field dependent core size effects. As for the KZ-finding that increasing $\lambda$ makes the system more BS-like, the $V_3Si$ data are consistent with this prediction: the $\rho_f/\rho_n$ curve is found to be closer to BSFF compared with $LuNi_2B_2C$, appearing to approach the KZ $t= 0.5$ curve at higher fields. The curves do not coincide even though both data sets are at approximately the same $t$. However, $V_3Si$ does have a higher scattering parameter $\lambda$; therefore the shift is consistent with the prediction of a weaker field dependence of the fluxon core size with higher $\lambda$, i. e., stronger scattering. It is interesting that the two curves might either merge or cross at lower fields (where the currents required to achieve FFF become difficult to attain), the reason for which is not yet clear and demands investigation by some other means.

## IV. CONCLUSIONS

Free flux flow resistivity levels determined by DC transport measurements in weakly-pinned systems are consistent with the Kogan-Zelezhina (KZ) model of a weak-coupling BCS manifestation of finite fluxon core size effects at low temperatures, in two different $s$-wave superconductors, $V_3Si$ and $LuNi_2B_2C$. Fig. 4 summarizes the main result. With a correction due to the known magnetoresistivity of $V_3Si$ in the normal state, the data appear to be consistent with the prediction that the field dependence of fluxon-core size is suppressed by both temperature and scattering. Finding such consistency with two different materials is remarkable, and underscores the value of performing a similar measurement using the same systems. In addition, this study also shows that free flux flow could be an insightful probe into properties of the fluxon core. Since fluxon core effects should also be detected with in-field specific heat[13], it would also be interesting to perform same-system specific heat measurements to see if similar consistency would be found.

## ACKNOWLEDGMENTS

This research was supported by an award from Research Corporation, along with undergraduate summer-research support from the University of South Alabama. Work at NHMFL is performed under the auspices of the State of Florida and the NSF under Cooperative Agreement No. DMR-0084173. Work at the Ames Laboratory was supported by the Department of Energy, Basic Energy

Sciences under Contract No. DE-AC02-07CH11358. Research at ORNL was supported by the U.S.D.O.E. Division of Materials Sciences and Engineering, Office of Basic Energy Sciences.

**FIGURE CAPTIONS:**

**Figure 1.** Temperature dependence of resistivity for the two compounds, showing critical temperatures $T_c$ (insets), and high residual resistivity ratios (RRR) that indicate low impurity levels. The inset graphs show closeups of the transition temperatures and are described in the Discussion of Results in the text.

**Figure 2.** Representative curves for the dissipation level, $V/I$, versus current for (a) LuNi$_2$B$_2$C at fields $H$ = 1.2, 1.5, 1.8, 2.0, 2.2, 2.5, 2.8, and 3.0 T; and (b) V$_3$Si at $H$ = 5.0, 5.5, …, 7.5, 8.0 T. Curves are plotted on semi-logarithmic axes to show saturation at highest currents and at field-dependent levels well below the normal-state dissipation level $R_n$ measured at $T$ = 4.2 K, $H = H_{c2}$ (values indicated by dashed line). Insets: Non-monotonic "peak effect" in the field dependence of critical current density $J_c$ at $T$ = 4.2 K, due to re-entrant weak pinning near $H_{c2}$ ; vertical line indicates upper bound for possibility of free flux flow (FFF), see text.

**Figure 3.** Temperature dependence of resistivity $\rho(T)$ measured at fields $H$ = 0, 1, 3, 4, 5, 6, 7, 8, and 9 T, plotted versus $T^2$. Martensitic transformation, MT, is marked, along with the measured $\rho_n$ ($T$ = 4.2 K, $H = H_{c2}$ = 18.3 T) = 1.56 µΩ. Open circles at $T$ = 4.2 K indicate magnetoresistivities obtained by extending the $T^2$ fits, which are then fitted with a Kohler-Rule form in the inset. See text for discussion.

**Figure 4.** Comparison of normalized flux-flow resistivity $\rho_f/\rho_n$ with predicted field dependence based on the Kogan-Zelezhina (KZ) model for $t = T/T_c$ = 0 and 0.5, along with traditional Bardeen-Stephen flux-flow (BSFF) model. At $t$ = (4.2 K)/$T_c$ , the data falls within these two curves. For V$_3$Si, black triangles show results after correcting $\rho_n$ ($H$) values for magnetoresistive effects; while uncorrected data using as-measured $\rho_n$ ($H_{c2}$) for all fields are shown as open circles; see text.

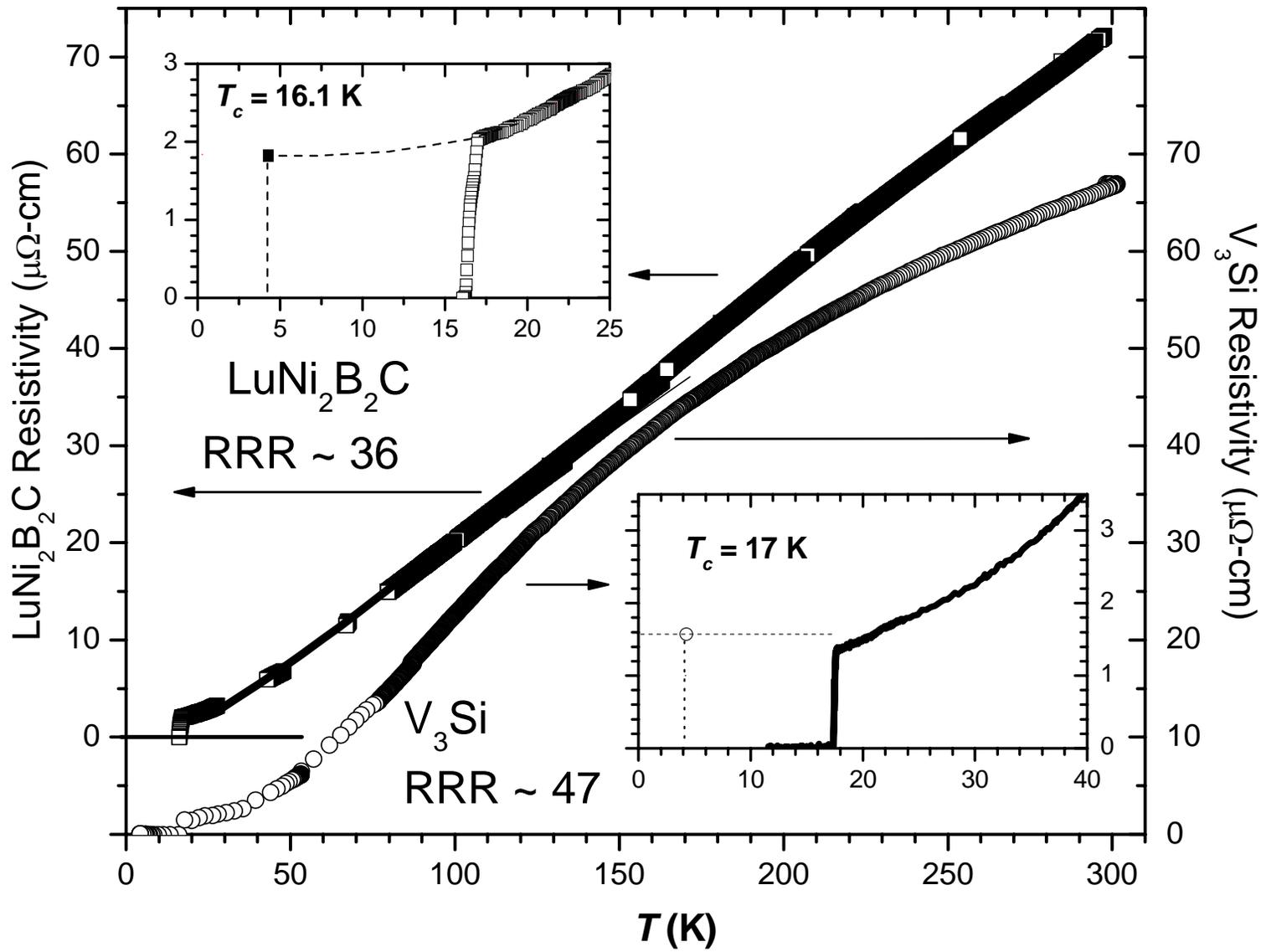

Figure 1 Gapud *et al.*

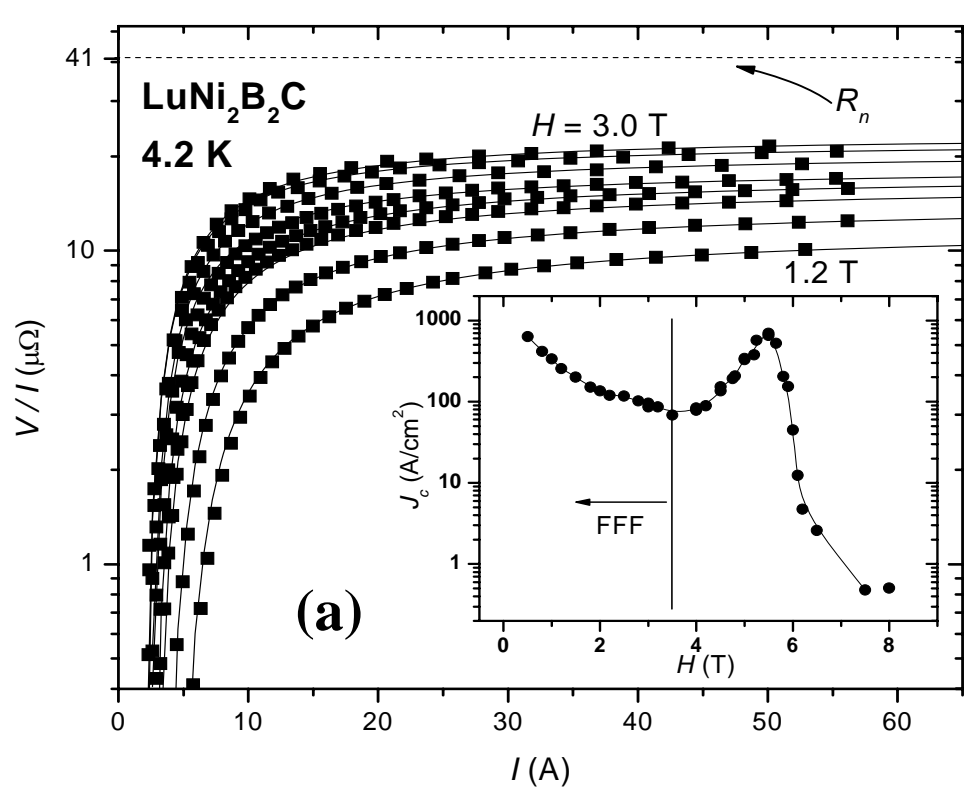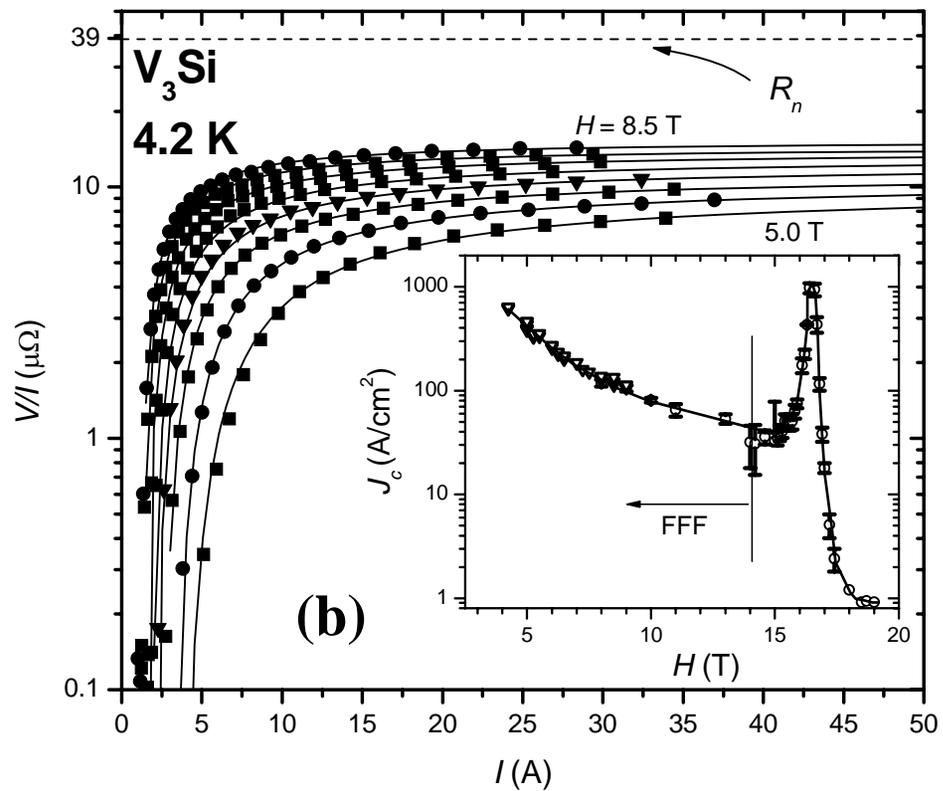

Figure 2 Gapud *et al.*

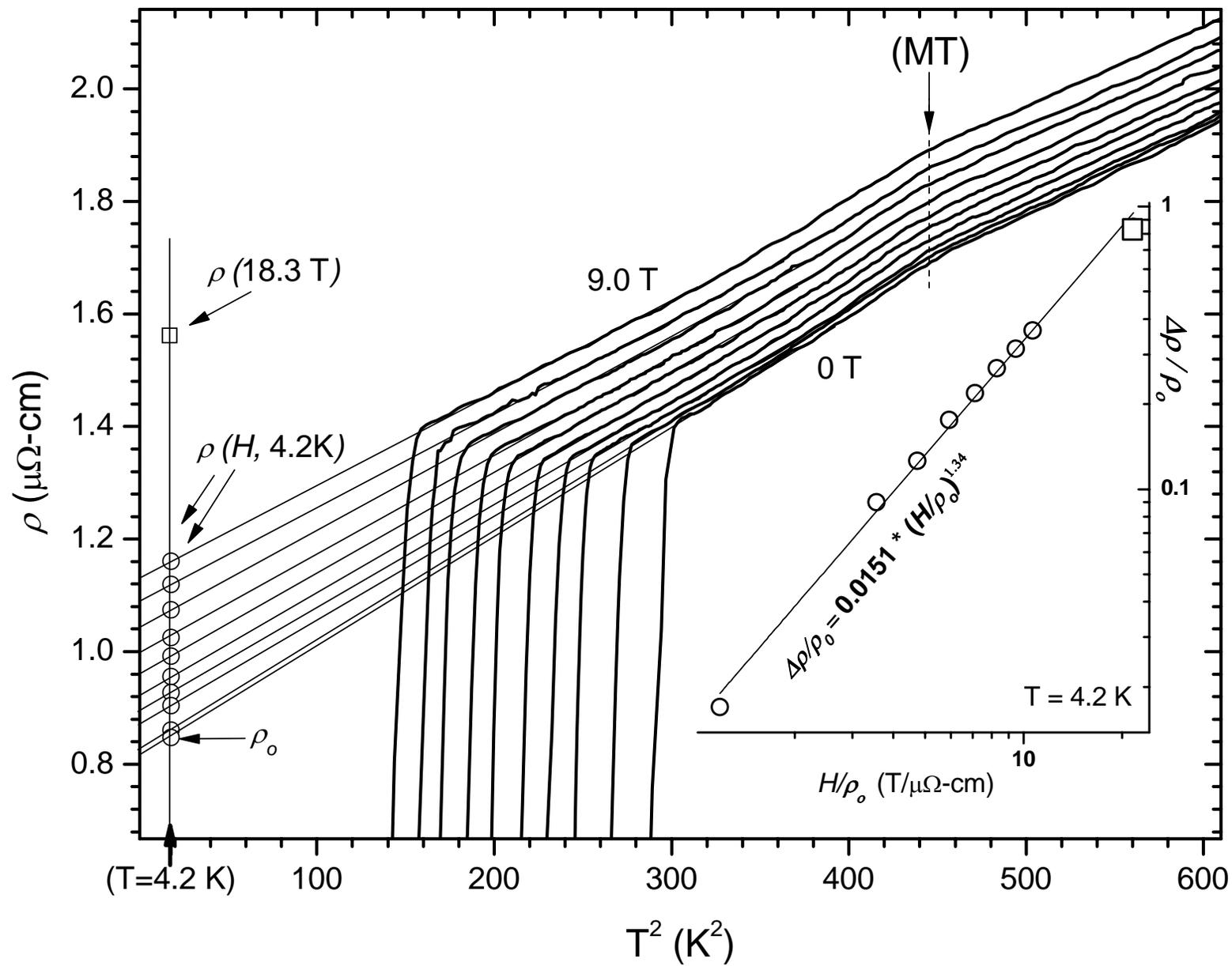

Figure 3 Gapud *et al.*

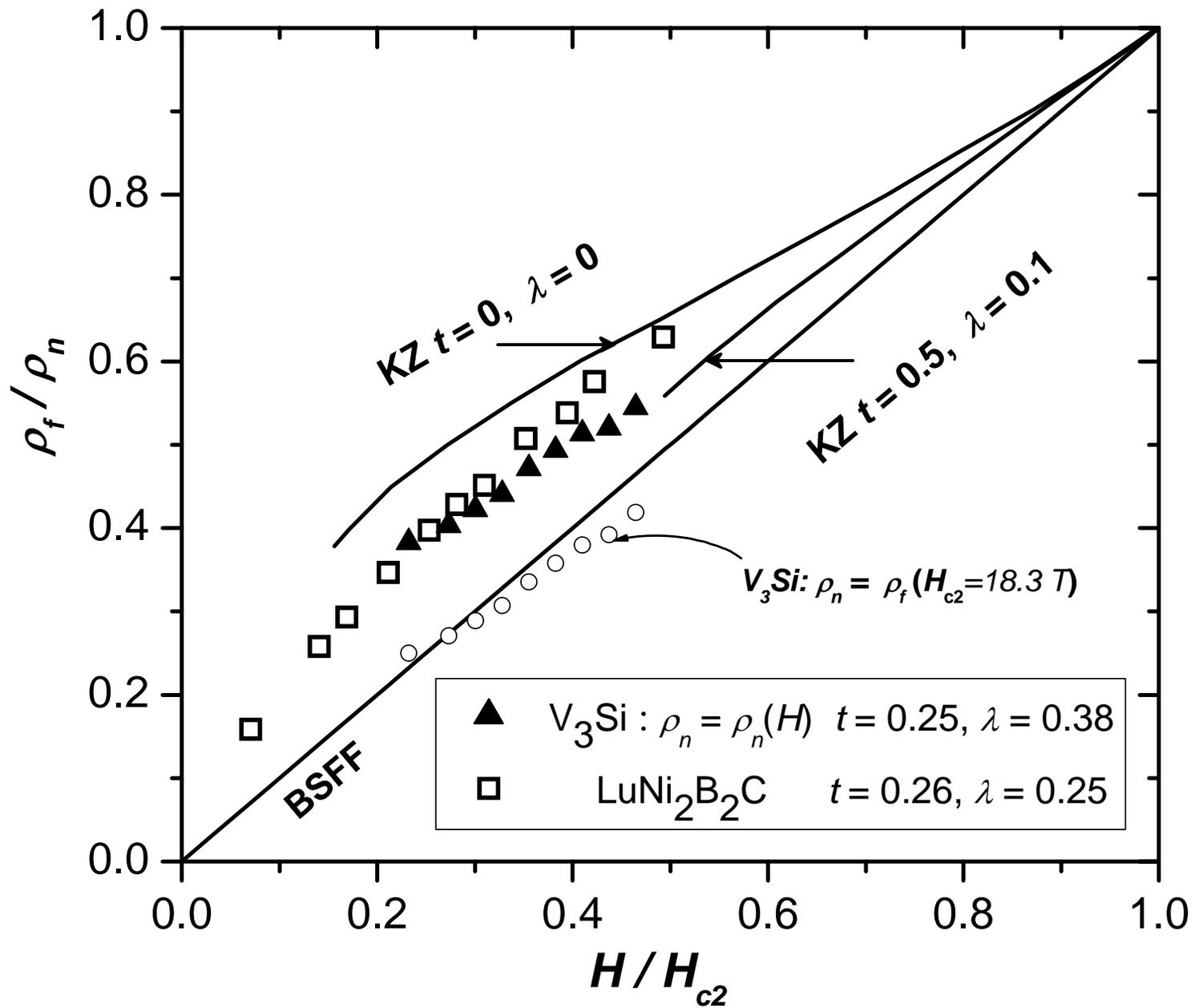

Figure 4 Gapud *et al.*